\documentclass[aps,floatfix,superscriptaddress,twocolumn]{revtex4-2}

\usepackage{mhchem}
\usepackage{multirow}
\usepackage{subcaption}
\usepackage{graphicx}
\usepackage{amsmath}

\begin{document}

\title{Accurate Crystal Structure Prediction of New 2D Hybrid Organic Inorganic Perovskites}

\author{Nima Karimitari}
\thanks{These authors contributed equally}
\affiliation{Department of Chemistry and Biochemistry, University of South Carolina, South Carolina 29208, United States}

\author{William J. Baldwin}
\email{wjb48@cam.ac.uk}
\affiliation{Department of Engineering, University of Cambridge, Cambridge CB2 1PZ, UK}

\author{\hspace{-4pt}$^{,\hspace{2pt}\ast}$ \text{Evan W. Muller}}
\affiliation{UES, Inc., Beavercreek, Ohio 45432, United States}

\author{Zachary J. L. Bare}
\affiliation{Department of Chemistry and Biochemistry, University of South Carolina, South Carolina 
29208, United States}

\author{W. Joshua Kennedy}
\affiliation{Materials and Manufacturing Directorate, Air Force Research Laboratory, Wright-Patterson AFB, Dayton, Ohio 45433, United States}

\author{G\'{a}bor Cs\'{a}nyi}
\affiliation{Department of Engineering, University of Cambridge, Cambridge CB2 1PZ, UK}

\author{Christopher Sutton}
\email{cs113@mailbox.sc.edu}

\affiliation{Department of Chemistry and Biochemistry, University of South Carolina, South Carolina 29208, United States}

\date{\today}

\begin{abstract}
Low dimensional hybrid organic-inorganic perovskites (HOIPs) represent a promising class of electronically active materials for both light absorption and emission. The design space of HOIPs is extremely large, since a diverse space of organic cations can be combined with different inorganic frameworks. This immense design space allows for tunable electronic and mechanical properties, but also necessitates the development of new tools for {\em in silico} high throughput analysis of candidate materials. In this work, we present an accurate, efficient, transferable and widely applicable machine learning interatomic potential (MLIP) for predicting the structure of new 2D HOIPs. Using the MACE architecture, an MLIP is trained on 86 diverse experimentally reported HOIP materials. The  MLIP is tested on 73 unseen perovskite compositions (that were previously reported experimentally), and achieves chemical accuracy with respect to the reference electronic structure method. Our model is then combined with a simple random structure search algorithm to predict the structure of new HOIPs given only the proposed composition as input. Success is demonstrated by correctly and reliably recovering the crystal structure of a set of experimentally known 2D perovskites. Such a random structure search is impossible with ab initio methods due to the associated computational cost, but is relatively inexpensive with the MLIP. Finally, the procedure is used to predict the structure formed by a new organic cation with no previously known corresponding perovskite. Laboratory synthesis of the new hybrid perovskite confirms the accuracy of our prediction using the combined MLIP and structure-search algorithm. This capability will enable the efficient and accurate screening of thousands of combinations of organic cations and inorganic layers for further investigation.
\end{abstract}

\maketitle

Hybrid organic-inorganic perovskites (HOIPs) belong to a broad category of materials, generally represented by the chemical formula \ce{ABX3}. The \ce{B}-site and \ce{X}-site ions form a network of corner-sharing \ce{BX6} octahedra. Although the \ce{A}-site can be a large inorganic cation, such as caesium, using an organic cation has proved extremely successful, resulting in the development of state of the art solution-processed optoelectronic materials \cite{hoip}. Provided the organic cation is small, the typical perovskite structure is retained. For larger cations, however, the network of corner sharing octahedra is disrupted, leading to `low dimensional' structures such as one-dimensional chains or two-dimensional sheets of octahedra (see Fig. \ref{fig:database_specs}b). 

Two-dimensional HOIPs are formed when the organic cations separate the inorganic layers in the (100), (110) or (111) direction, giving the modified general formula \ce{{A'}_{m}A_{n-1}B_{n}X_{3n+1}}. The constants $n$ and $m$ determine the number of connected inorganic layers and the charge of the organic cation. They are further categorized into two main types: Dion–Jacobson (DJ) \cite{dj} with $m=1$ (one sheet of interlayer cations with +2 charge) and Ruddlesden–Popper (RP) \cite{rp1,rp2} with $m=2$ (two sheets of cations with +1 charge) \cite{2d_1}. Two dimensional HOIPs have the advantages of enhanced stability under ambient conditions and structural tunability. This makes them promising candidates for applications in photoluminescence (PL), photovoltaics, photodetection, and light emmitting diodes (LEDs) \cite{app1,app2,app3,app4}.

Due to the breadth of the design space of 2D (as well as 1D and 0D) perovskites, {\em in silico} property screening is desirable. However, in order to calculate properties with ab initio electronic structure methods, one first needs to know the crystal structure. A similar task has been tackled in the field of organic crystal structure prediction (CSP): typically, CSP methods involve generating many hundreds or thousands of candidate structures, and selecting the lowest energy structures using an empirical force field \cite{Day2011}. For general inorganic crystals, the related Random Structure Search (RSS) method has been successful for unit cells of up to a couple of dozen atoms, wherein candidate structures are generated and the geometry is subsequently relaxed to the nearest local minima in the potential energy landscape \cite{Pickard_2011}. 

A particular difficulty in the case of 2D HOIPs is that they can have extremely large unit cells containing up to 1000 atoms. Furthermore, they are structurally complex (see Fig.~\ref{fig:database_specs}) with the organic molecules having many potentially quite flexible degrees of freedom. Direct Density Functional Theory (DFT) geometry relaxations or molecular dynamics simulations are therefore extremely expensive, while empirical force fields which are accurate across the desired range of chemical interactions do not exist presently. 

An alternative to doing ab initio calculations is to use machine-learned interatomic potentials (MLIPs) \cite{Bartok2017, Musil2021, Butler2018, Deringer2021}. MLIPs can be trained to predict the potential energy of a configuration of atoms directly from the atomic coordinates, allowing for simulations of hundreds of thousands of atoms at DFT accuracy \cite{Deringer2021silicon}. Many MLIP architectures have been developed in recent years. Key developments in this area have been the focus on atom-centered energy contributions enabling linear scaling models, the incorporation of physical symmetries into model architectures \cite{Behler2007GeneralizedSurfaces,Bartok2010GaussianElectrons, BartokSOAP2013} and efficient construction of many-body representations of atomic environments \cite{Shapeev2016MTP, ace1, Drautz2020}. Furthermore, the introduction of graph models to MLIP development has lead to greatly improved accuracy and transferability \cite{MPNNGilmer2017, Batzner2022:nequip, Batatia2022ThePotentials}. MLIPs have already been used to perform structure prediction for large scale screening tasks, for example in a computational study searching for novel stable inorganic materials \cite{Merchant2023}.

In this work, the MACE \cite{MACE2022} message passing architecture was used to build a transferable MLIP for HOIPs. MACE is a graph tensor network which constructs many-body equivariant messages at each node (nodes correspond to atoms in this case) via the atomic cluster expansion \cite{ace1}, which are then passed onto neighbouring nodes. The architecture has been shown to be accurate, efficient and transferable \cite{KovacsMACEeval2023}, and has recently been used to create a state of the art ML organic force field\cite{maceoff23} and a ``foundation model'' for materials chemistry\cite{batatia2024foundation}. The model in this work is fitted to data collected from several publicly available databases of experimentally reported HOIPs. Starting from structures reported in these databases, an extensive training dataset was generated by running an active learning protocol based on molecular dynamics. Collected configurations were labelled with DFT calculations. The final model achieves excellent accuracy across an independent set of perovskites with unseen compositions taken from the same sources. 

To use the model effectively, we present a simple random structure search procedure designed for 2D HOIPs and we show that the trained MLIP accurately captures the complex potential energy landscape encountered during a random structure search task. Furthermore, the combination of the structure searching algorithm and the MACE model is an accurate and efficient structure prediction tool. This is shown by `re-discovering' the ground state structure of a set of experimentally reported HOIPs not seen by the model during fitting, given only the most basic information of the perovskite - the identity of the organic cation and the composition of the inorganic layer. 

Finally, we predict the crystal structure of a previously unknown 2D HOIP. We then synthesize the material in the laboratory, and verify that the structure agrees with our prediction. The process reveals a large number of competing low energy minima, with subtly different orientations and stacking patterns of the organic cation. Due to the high degree of similarity between these structures, an accurate and efficient search tool offers many insights beyond just prediction of the ground state structure. 

\section{Dataset Construction}

A dataset was compiled from three sources: The 2D perovskites database of the laboratory of new materials for solar energy (NMSE) \cite{nmse}, the Cambridge Structural Database \cite{CCDC}, and a recent research article by Tremblay et al. reporting numerous 2D HOIP structures \cite{tremblay2022hybrid}. 

\begin{figure*}[htb]
    \centering
    \includegraphics[width=1.0\textwidth]{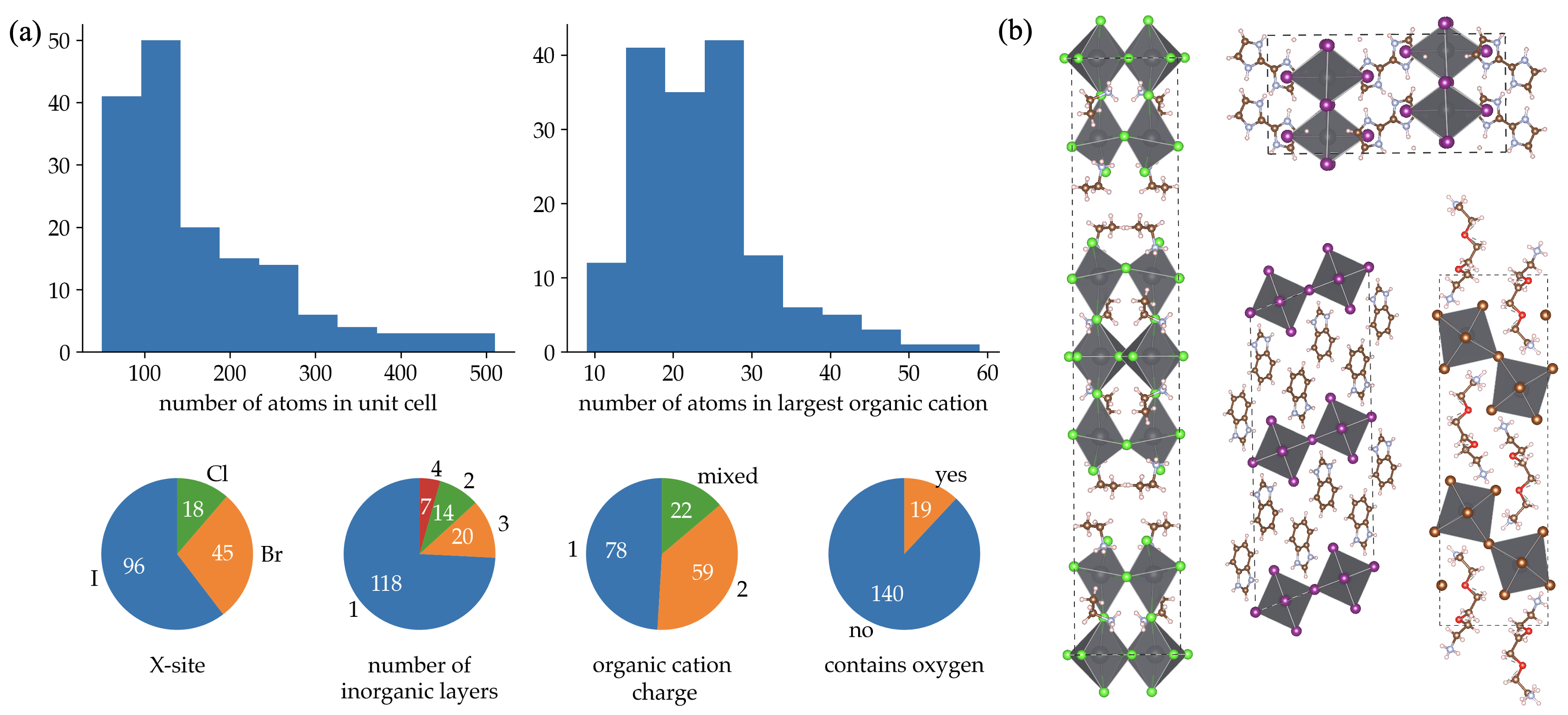}
    \caption{(a): Key properties of the perovskites in the compiled dataset. Note that some structures have multiple organic cations, but the upper right histogram shows only the size of the largest cation in each structure. (b): Examples of 2D HOIP structures in the dataset.}
    \label{fig:database_specs}
\end{figure*}

The occurrence of different chemical elements and structural features in these sources was quite non-uniform. Several simplifying restrictions have therefore been placed on the scope of our model. Firstly, the set of chemical elements considered for the inorganic layer was restricted to only include \ce{Pb, I, Br} and \ce{Cl}. As a result, the resulting MLIP can be applied to only Pb-based perovskites, with X = I, Br, or Cl. Furthermore, we restricted the composition of the organic cation to include only C, H, N and O. These restrictions were imposed due to the occurrence of different chemical elements in the available 2D HOIP datasets: of the structures we collected, more than 80\% were lead-based, and the majority contained only C, H, and N elements in the organic cation. Applying these filters resulted in an initial dataset of 159 experimentally reported structures. Fig. \ref{fig:database_specs}a presents some key statistics of this dataset including the number of atoms in the unit cell and organic cation, as well as a breakdown of the elements present at the X-site, number of inorganic layers, organic cation charge and whether the organic cation contains oxygen. Four representative example structures are shown in Fig.~\ref{fig:database_specs}b to illustrate the diversity of the perovskites that are included. 

\section{Model Development and Performance}

\subsection{Active Learning for Dataset Expansion}

In the following, \textit{composition} will refer to a perovskite as determined by the chemical formula and the experimentally reported unit cell, while \textit{configuration} will refer to a specific non-equilibrium set of atomic positions, for which one could compute a reference energy using DFT. The dataset described above serves as a starting point for fitting a MLIP. In practice, however, fitting accurate and stable models requires a database with many non-equilibrium configurations for each target composition or phase. One popular method for database construction is to sample configurations from molecular dynamics trajectories. In this study, a different approach is taken in which a database of reference configurations is grown iteratively in an active learning procedure\cite{Behler_how_to_train, vanderOord2022}.

Before beginning the active learning procedure, the dataset of experimentally reported compositions was first divided in two, by randomly sampling 86 perovskites to form the core of the training set. The remaining test set compositions will be used to assess the transferability of the final model to new unseen perovskites.

The key principle of active learning is to use a model which can estimate the uncertainty of its own prediction on a given configuration. If this estimate is reliable, one can search for configurations on which the model is uncertain, and add only these configurations to the dataset. Several methods exist for constructing MLIPs with an in-built measure of prediction uncertainty. For MACE, the uncertainty estimate can be obtained as follows: given a dataset, one fits several independent models with the same hyperparameters, but with a different random initialisation of weights. These models are then referred to as a committee, and in this case, we use only 3 models to form the committee. On a new configuration, the disagreement between the committee members can be treated as an uncertainty estimate. As will be shown below, this comparatively inexpensive procedure leads to remarkably useful uncertainty estimates. 

With this method for assessing the uncertainty of a model, the active learning procedure is as follows:
\begin{enumerate}
    \item Given an initial dataset of configurations, calculate reference energies and atomic forces using DFT. Fit a committee of 3 MACE models on this dataset.
    \item For each composition in the training set, run an MD simulation starting from the experimentally reported structure and using the average of the force predictions of the committee members to propagate the dynamics. At each time step, test the uncertainty of the potential by calculating the disagreement in the prediction of the atomic forces between the committee members.
    \item If the {\em relative} force uncertainty of any atom, defined as the standard deviation of the committee force predictions divided by the mean of the forces, is larger than a specific threshold (in our study this threshold is to 0.2, see also section \ref{sec:methods_md}) the MD simulation is terminated. DFT energy and forces are calculated for the configuration for which the uncertainty exceeded the threshold, and added to the training set. If the uncertainty does not exceed the threshold within  10~ps, terminate the MD and do not collect any new configurations. 
    \item Refit the committee of models with the expanded dataset. It is expected that the configurations where the models previously disagreed are now well described with low uncertainty. 
    \item Repeat steps 2-5 until no new configurations are collected for any of the compositions. 
\end{enumerate}

In each cycle of active learning, we ran committee MD for each of the 86 compositions in the training set. Additional configurations are therefore collected at a rate of 86 per cycle if the uncertainty for all compositions exceed the threshold. However, as the dataset grows, many compositions quickly become well described and do not trigger new DFT calculations, resulting in few new training configurations per cycle. For this reason, in later cycles step 3 was repeated 10 times for each training set composition before retraining the model. The final potential is fitted once all the unique perovskite compositions in the training set are stable, meaning that in 10 independent MD simulations lasting 10~ps each, the 0.2 relative force uncertainty threshold is not exceeded. Additional details on the active learning procedure are given in section \ref{sec:methods_md}

\begin{figure}
    \centering
    \includegraphics[width=\linewidth]{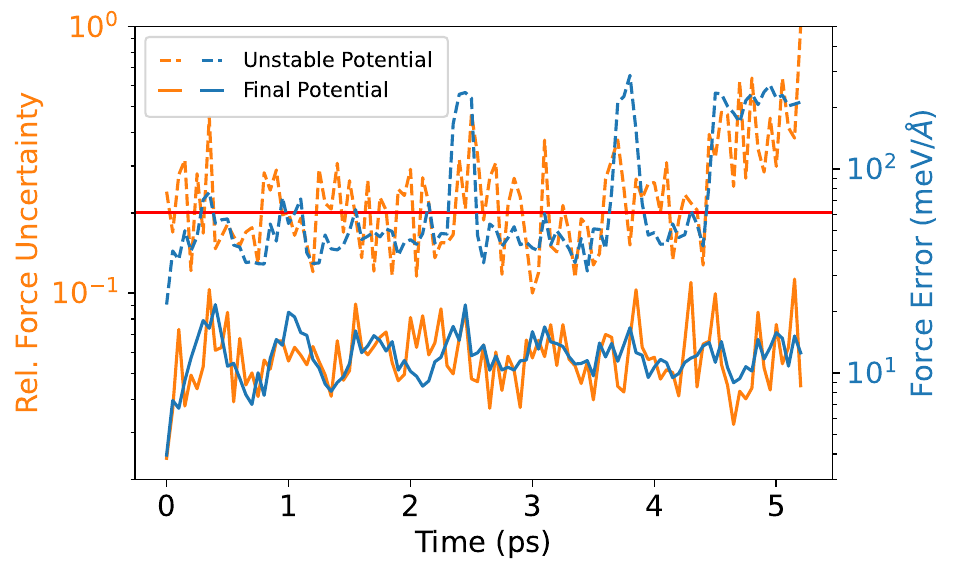}
    \caption{ The relative force uncertainty and actual force error for one HOIP as a function of time during an MD simulation. The unstable potential (dashed lines) occasionally exceeds the relative force uncertainty threshold (red solid line at 0.2) with actual force errors as large as 100~meV/\AA, while the final potential remains far below the threshold with force errors fluctuating between 10 to 20~meV/\AA.}
    \label{fig:unc}
\end{figure}

Key to this method is the reliability of the uncertainty measure. An example of the evolution of the uncertainty measure during a committee MD simulation is shown in Fig. ~\ref{fig:unc}. To assess the uncertainty measure, all configurations in the trajectory were evaluated with DFT, and the actual force error made by the model at each time step is also shown. Two models from different stages in the active learning procedure are shown - an `unstable' model from an early point in the active learning process, and the the final model.

For the unstable potential, the uncertainty exceeds the threshold (the solid red line) at multiple instances, and eventually increases to 1.0 implying total uncertainty in force predictions. By contrast, the final potential has both a consistently lower uncertainty and a lower force error. The key result shown in Fig.~\ref{fig:unc} is that the difference in force error between the final model and unstable model is clearly reflected in the estimated uncertainty. Also important is that the spikes in the force error of the unstable model closely correlate with the spikes of the relative force uncertainty.

In general, the highest uncertainty occurs for atomic configurations that are less represented in the training set. In particular, there are 61 unique organic cations in the 86 compositions of the training dataset, while there are just 7 types of inorganic layer. Therefore, the highest force uncertainty typically occurred on the organic cations. 

\subsection{Final Model Performance}

\begin{figure}
    \centering
    \includegraphics[width=0.9\linewidth]{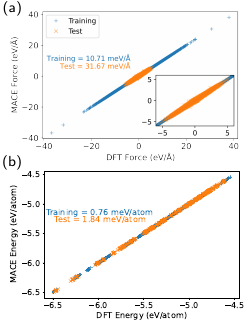}
    \caption{The parity plot of (a) forces, and (b) energy (per atom) for training and test set samples.}
    \label{fig:energy}
\end{figure}

In total, 18 cycles of active learning were performed. The final training dataset contains a total of 2457 configurations. To test the final potential, MD simulations of 73 unseen test set compositions were ran for 10 ps and samples were taken every 1 ps. The energy and force predictions for all the training and test samples are shown in Fig.~\ref{fig:energy}. The RMSE of training (test) dataset for energy and forces are 0.76 (1.84) meV/atom and 10.7 (31.7) meV/\AA, respectively. In addition, the errors categorized based on the halide atoms are shown in Table.~\ref{table:comparison}. An energy error of 1 kcal/mol (typically called ``chemical accuracy'') corresponds to 43.4 meV per formula unit. 

\begin{table}
\centering
\caption{\label{table:comparison} Energy and force errors for seen and unseen configurations categorized based on the halides}
\begin{tabular}{p{1.2cm} p{1.7cm}  p{1.7cm}  p{1.7cm} p{1.7cm}}
 & \multicolumn{2}{c}{Seen Compositions} & \multicolumn{2}{c}{Unseen Compositions}\\
\hline
\hline
 & Energy (meV/atom) & Forces (meV/\AA)  & Energy (meV/atom) & Forces (meV/\AA) \\   
\hline
 Cl & 0.86  & 9.25 & 1.86 & 30.4  \\
 I  & 0.74  & 10.96 & 1.34 & 29.88  \\
 Br & 0.78 & 10.39 & 2.12  & 48.53  \\ 
 \hline
 Total & 0.76 & 10.71 & 1.84  & 31.67  \\ 
 \end{tabular}
 \end{table}

\section{Relaxation of Experimentally Reported Structures}

Experimentally reported structures are typically close to the global minima of the potential energy surface. For the trained MACE model to be useful for structure searching, it must relax these structures to the same local minima as would be obtained by a DFT geometry relaxation. To assess whether this is the case, we considered 137 perovskite compositions in the dataset that have less than 200 atoms, with 58 from the training set and 79 from the test set. For all of these compositions, the experimentally reported structure was relaxed independently with DFT and MACE calculators, until the forces were less than 10 meV/\AA.

\begin{figure*}
    \centering
    \includegraphics[width=1.0\textwidth]{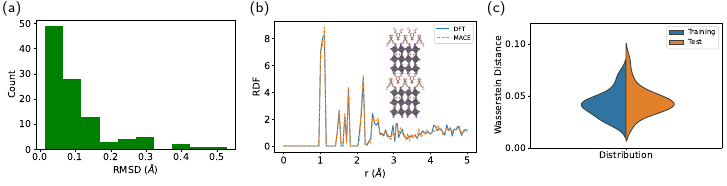}
    \caption{Evaluating the MACE model for geometry relaxations of experimentally reported structures.  (a) Histogram of the RMSD between the DFT relaxed and the MACE relaxed structures for the entire dataset. (b) Comparison of the total RDF for a test set structure after relaxing with DFT and MACE. (c) The distribution of the Wasserstein distance between the RDFs given by the DFT relaxed structure and MACE relaxed structure for 137 unique HOIPs in the training and test set.}
    \label{fig:DFT_MACE_relaxations_comparisons}
\end{figure*}

One way to quantify the difference between the MACE and DFT relaxed structures is to measure the root mean square displacement (RMSD) of the atoms between the two structures. The distribution of RMSD for all 137 compositions is shown in Fig.~\ref{fig:DFT_MACE_relaxations_comparisons}a. For the majority of the samples, the RMSD is less than 0.1~\AA. Several outliers are present with larger RMSDs on the order of 0.3-0.5~\AA. These outliers generally correspond to cases in which long, flexible organic molecules move slightly with respect to each other. 

The independently obtained DFT and MLIP relaxed structures can also be compared using the total radial distribution function (RDF), which contains information about the bond lengths, intermolecular distances and organic-inorganic distances in the structure. A comparison between the RDFs of a MACE and DFT relaxed structure is shown in Fig.~\ref{fig:DFT_MACE_relaxations_comparisons}b. For $r<3$~\AA, which mostly corresponds to the intramolecular bond distances, the differences between MACE and DFT are negligible. For $r>3$~\AA, which contains both the intermolecular distances and inorganic bonds, some differences are apparent, however the structures relaxed with MACE and DFT share many of the larger features.

To quantify the difference between the RDFs of MACE and DFT relaxed structures, we used the first Wasserstein distance (the earth mover's distance, EMD) between these two distributions, which calculates the least amount work required to change one distribution to the other \cite{EMD}. A histogram of the Wasserstein distances for 63 randomly selected compositions in the train and test sets is shown in Fig. ~\ref{fig:DFT_MACE_relaxations_comparisons}c. One can see that the final MLIP performs similarly for both training and test sets using this metric.

\section{High Throughput Structure Prediction for New HOIPs}

Calculating properties of known perovskite structures with ab initio methods is expensive, but not impossible. On the other hand, high throughput structure prediction for many new compositions is potentially an infeasibly expensive task, particularly for structures with large unit cells. This is because crystal structure prediction protocols typically involve a very large number of either geometry relaxations or single point evaluations to predict the structure of just one chemical composition. 

In particular, organic crystal structure prediction involves first generating many (thousands) of candidate crystal structures by enumerating over key variables, such as space groups, and employing heuristics. Single point evaluations with empirical force fields are used to select good candidates, based on lowest potential energy\cite{Day2011}. 

Ab initio random structure search (AIRSS), is another approach that has been explored \cite{Pickard_2011}, particularly for inorganic crystals. In this approach, crystal structures are determined by first guessing random positions of atoms within the unit cell, followed by geometry relaxations with DFT. Again, the lowest energy structure is chosen as the most probable structure. AIRSS has been employed successfully to find ground state structures of materials, molecules and features such as defects \cite{Pickard_2011}. This process is powerful, but limited to small unit cells due to the poor scaling of DFT. 

In the following we introduce a simple structure search procedure inspired by these ideas, which is appropriate for 2D HOIPs.

\subsection{A Random Structure Search Procedure for 2D HOIPs}

\begin{figure}
    \centering
    \includegraphics[width=0.9\linewidth]{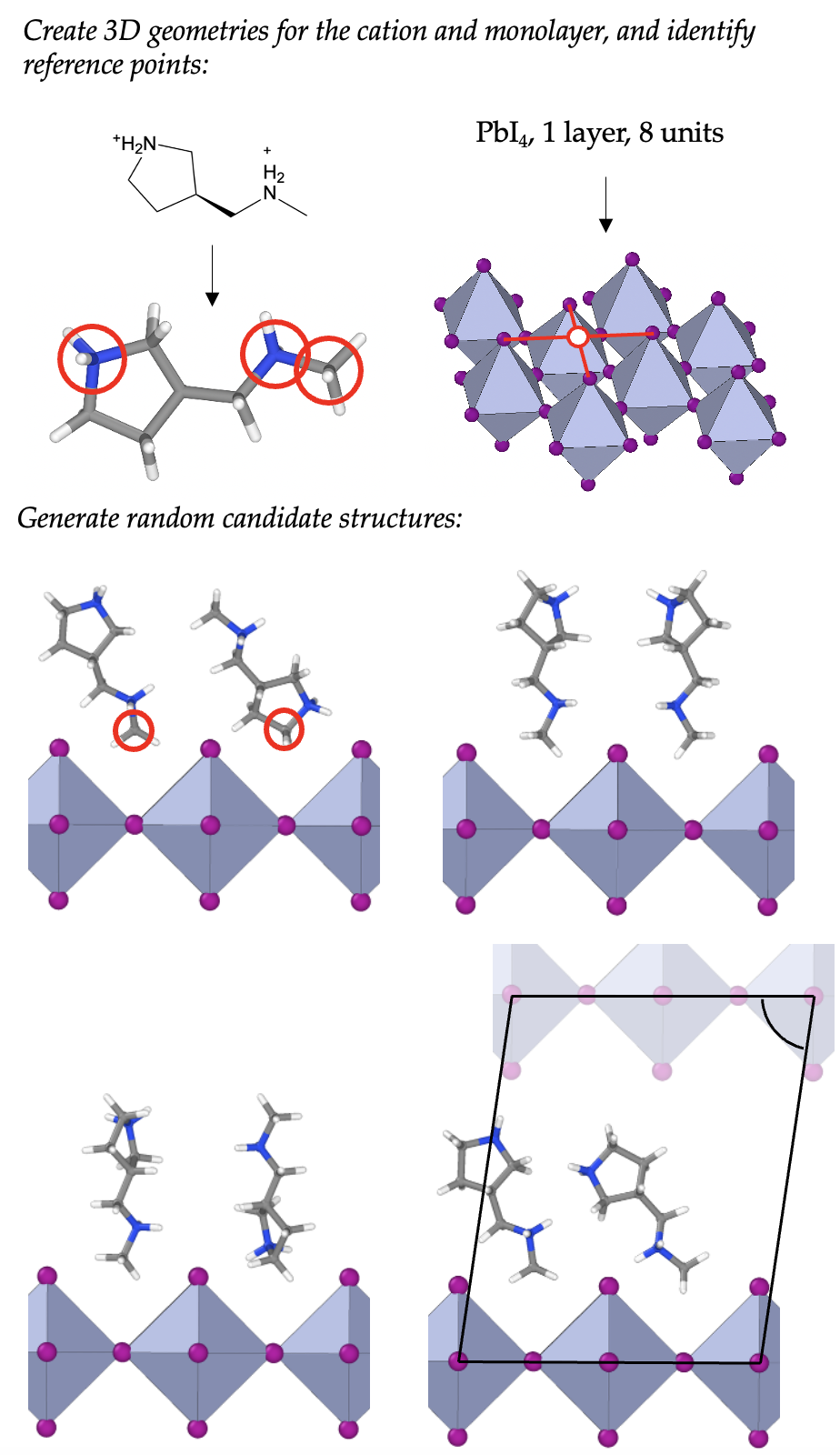}
    \caption{Overview of the structure generation algorithm for creating initial guesses for the random structure search process. To make the figure more readable, the unit cell is only shown for one of the four candidate structures in the lower panel.}
    \label{fig:random_generation_process}
\end{figure}

Our proposed structure searching workflow is summarised as follows: For a given organic cation and inorganic layer, generate a fixed number of candidate structures, which cover the space of feasible molecular and atomic arrangements. The geometry of all structures is then relaxed to a local minimum using the MLIP and the lowest potential energy structure is declared as the most probable crystal structure. The process for generating random candidate structures is key, and a scheme was designed based on several simple heuristics. The steps are summarised as follows and shown visually in Fig. \ref{fig:random_generation_process}:
\begin{enumerate}
    \item The starting information is the identity of the organic cation, the choice of halide, and the desired size of the unit cell. The size is determined by the number of organic/inorganic layers, and the number of octahedra per layer in the unit cell. 
    \item For the given composition, construct the 3D geometry of the organic cation (enumerating or sampling conformers if necessary). Also construct the untilted, strain-free inorganic layer from lead and the chosen halide. This determines the periodicity of the system in the in-plane directions. 
    \item Identify `reference points' on the cation and the inorganic layer. On the cation, reference points are defined as formally charged atoms or salient atoms. On the inorganic layer, the reference points are chosen to be the midpoints between protruding halides, as shown in Fig. \ref{fig:random_generation_process}.
    \item Based on the charge of the organic cation, determine the number of cations per layer required for charge neutrality. For each organic cation in the unit cell, randomly generate a set of reflections and rotations to apply to the organic cation. Subsequently, place the transformed cations onto the inorganic layer by pairing reference points on the two geometries. 
    \item Check for any intersections between cations, or intersections of cations with the inorganic layer. Discard samples for which these components intersect one another. 
    \item Fix the lattice constant in the out-of-plane direction to remove most of the vacuum region from the cell, including some amount of shear of the unit cell. If more than one inorganic/organic layer per unit cell is desired, repeat the above procedure and stack the resulting geometries. 
\end{enumerate}

This process gives structures which sample the configuration space well but which can contain high energy features, such as regions of vacuum or atoms at energetically unfavourable separations. Crucially, the configurations are sufficiently sensible that geometry relaxation leads to reasonable structures. 

A python package was written to implement this algorithm which is available at \url{https://github.com/WillBaldwin0/LDHP-builder}. The algorithm is specific to 2D corner-sharing HOIPs, since it relies on heuristics when placing molecules onto the inorganic layer. In practice, it was found that these heuristics perform remarkably well. Further details are provided in section \ref{section:methods_random_algo}.

\subsection{Validation of the Model on Randomly Generated Structures}

For an MLIP to be useful for the structure prediction task, the model must be accurate for the randomly generated structures and must not exhibit many spurious local minima. Crucially, it should reliably relax the structures to nearby DFT minima. 

\begin{figure*}
    \centering
    \includegraphics[width=1.0\textwidth]{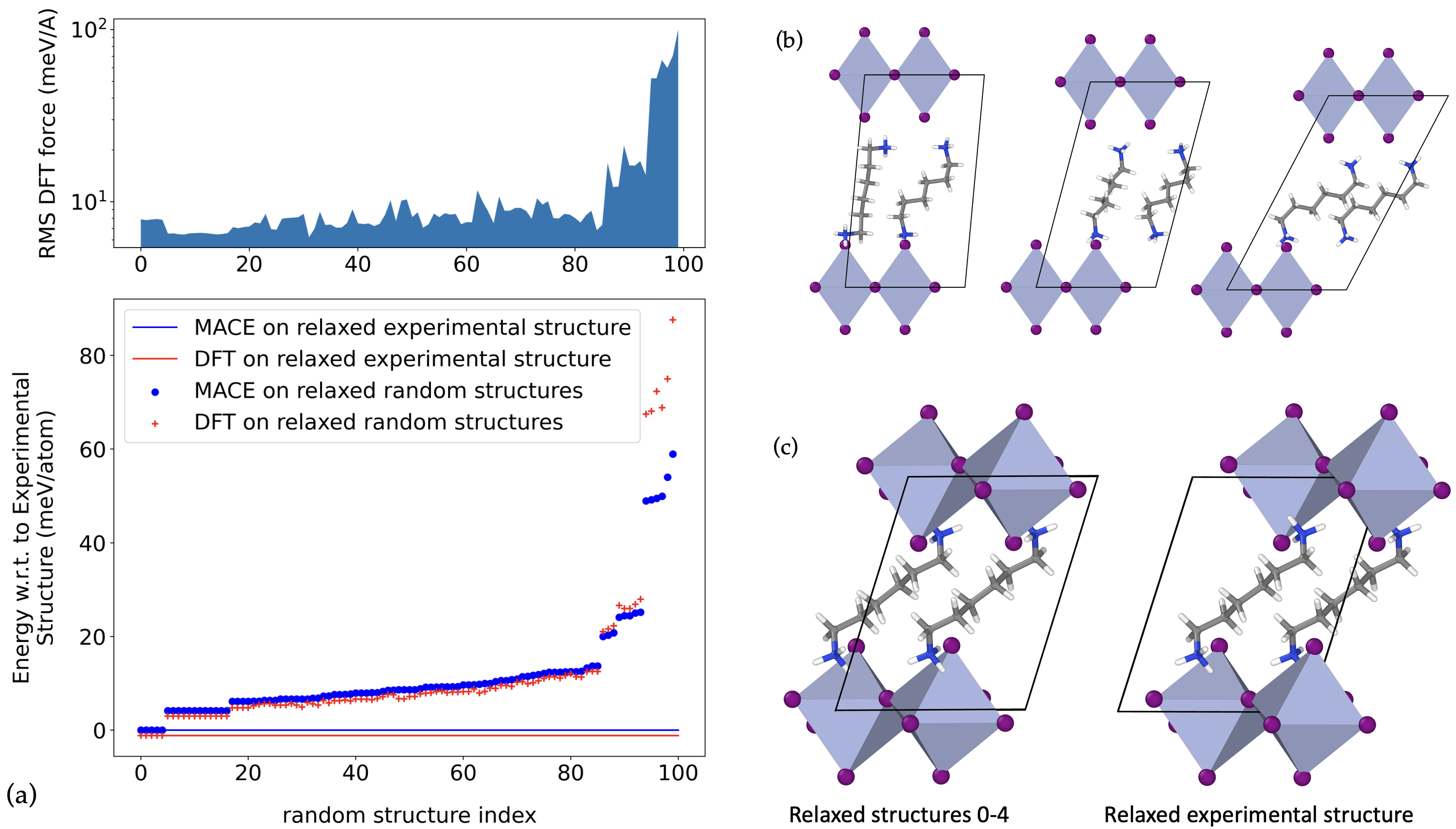}
    \caption{Rediscovering the structure of known a 2D perovskite. (a) Lower panel: formation energy (meV/atom) (blue) for the 100 randomly generated candidate structures, after geometry relaxation with the MLIP. Structures have been ordered according to increasing energy. Red points and lines show the same structures re-calculated with DFT. Upper panel: Root mean square forces, according to DFT, of each of the relaxed structures. (b) Examples of the initial random configurations. (c) Comparison between the structure obtained by relaxing the experimentally determined structure, and the five lowest energy structures found by the screening method.}
    \label{fig:random_relaxation_example}
\end{figure*}

To demonstrate the accuracy of the model and structure searching method, we present the results of the process applied to a known 2D perovskite in Fig.~\ref{fig:random_relaxation_example}. Specifically, we take the the perovskite formed by \ce{PbI6} octahedra in the inorganic layer and the organic cation \ce{NH3+[C]6NH3+}. The (geometry relaxed) experimentally reported structure is shown on the right of  Fig.~\ref{fig:random_relaxation_example}(c). 

Given the composition, 100 random structures were generated using the random generation procedure. Three examples of such structures are shown in Fig. ~\ref{fig:random_relaxation_example}b. To simplify this demonstration, only the correct conformer of the organic cation was used to generate samples. Subsequently, these 100 structures were relaxed using the MACE model. Since the initial samples are relatively high in energy - often containing a considerable amount of vacuum or non-physical molecular arrangements - these geometry relaxations require several hundreds or even thousands of steps. Fig. \ref{fig:random_relaxation_example}(a) shows the distribution of energy of the resulting structures, ordered by increasing energy, relative to that of the experimentally reported structure. Also shown is the energy of the relaxed samples after re-evaluation with DFT.

Several important features can be noted. Firstly, due to the nature of the long organic cation, which can stack in a variety of ways, the relaxation process reveals many local minima in the potential energy landscape. These appear as plateaus in the energy plot (bottom panel of Fig. \ref{fig:random_relaxation_example}(a)). After re-calculation of these structures with DFT, we see that the MLIP energy landscape is broadly correct in that these minima are correctly ordered with respect to DFT. The absolute energy error is also very low, being around 1 meV/atom which is roughly the accuracy of the model. Furthermore, the top of panel of Fig. ~\ref{fig:random_relaxation_example}(a) shows the root mean square force, according to DFT, of the MLIP identified minima. For all but the highest energy configurations, the DFT forces are less than 10 meV/\AA, suggesting these are close to DFT minima. 

The lowest energy structures identified by this procedure (the first five blue marks in Fig. \ref{fig:random_relaxation_example}a) have energy equal to that of the experimentally reported structure. This suggests that the process has indeed re-discovered the the experimentally reported structure. This was confirmed by examining the five lowest energy relaxed structures. Up to rotations, reflections and cell reductions, these structures are identical and match the experimentally reported structure as shown in Fig. \ref{fig:random_relaxation_example}(c).

\subsection{Structure Prediction Performance across the Dataset}

\begin{figure*}
    \centering
    \includegraphics[width=1.0\textwidth]{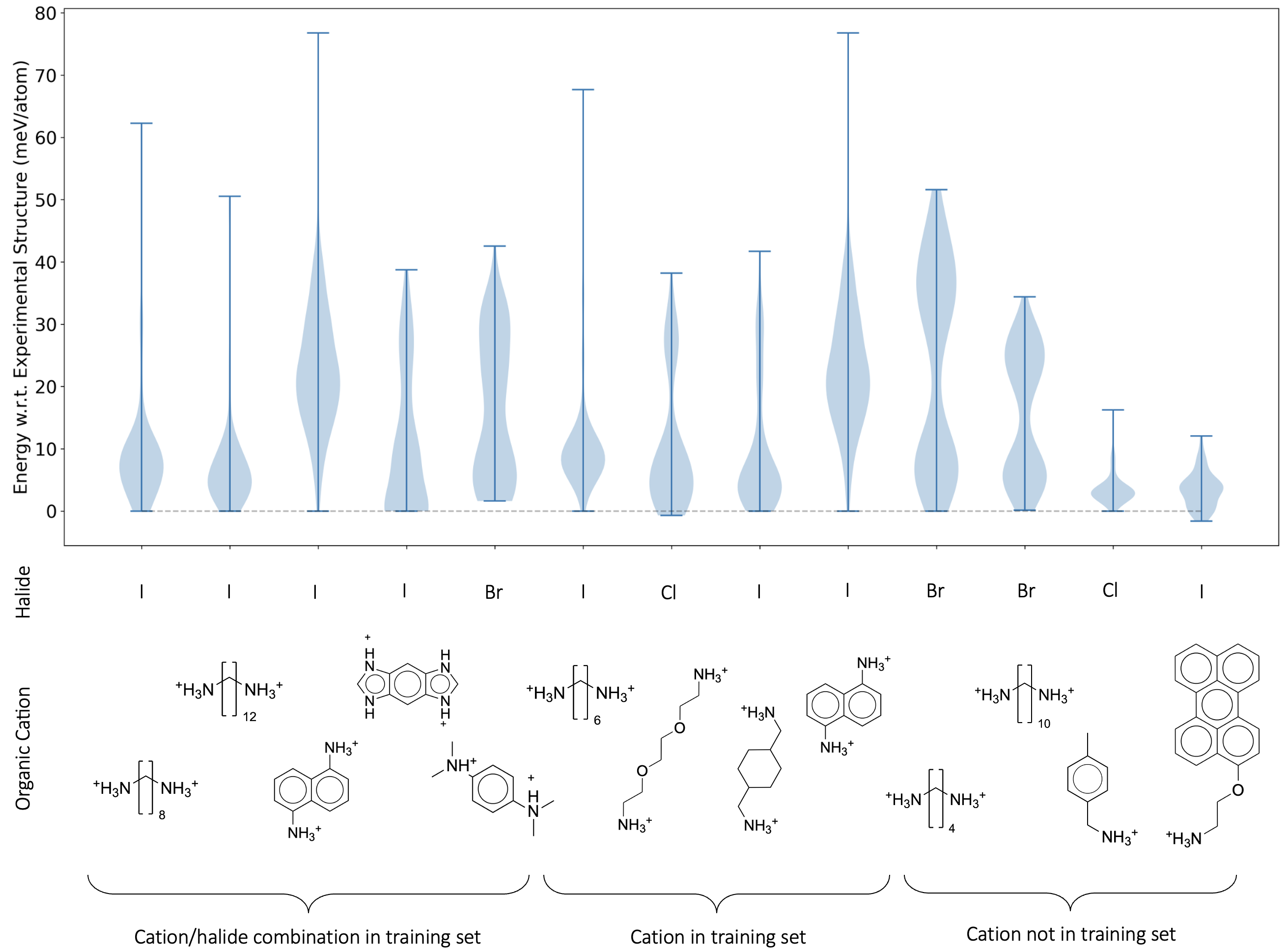}
    \caption{Performance of the random structure searching protocol applied to 13 experimentally known structures. Lower panel: Each combination of halide and organic cation shown in the lower part of the figure describes a perovskite present in the dataset. Some of these structures were used to train the model whereas others are unseen. Upper panel: Violin plots of the energy distribution of the random samples after geometry relaxation with MACE, relative to the energy of the geometry relaxed experimental structure. In the ideal case, the lower end of each violin plot would sit on the dashed line, indicating that the minimum energy structure found by the procedure is indeed the experimentally reported one. Structures are grouped into three categories. The first group contains perovskites which are present in the training set of the model. Following this are structures for which the combination of halide and cation is not present in the training set, but the cation is present paired with a different halide. The last group contains structures where the organic cation is not present anywhere in the training set.}
    \label{fig:random_structure_prediction}
\end{figure*}

We now demonstrate the usefulness of this procedure across a wider variety of 2D perovskites. The method described above has been applied to 13 structures in the dataset. Fig.~\ref{fig:random_structure_prediction}
summarises the results of this process. The lower rows identify the perovskite structure, via the halide in the inorganic layer and the organic cation. The upper panel shows the distribution of energies of the relaxed structures, following the random generation and relaxation process, with respect to that of the experimentally known structures. For this demonstration, 200 random structures were generated for each halide/cation combination. Only 200 samples were required, since all but the last two structures in figure \ref{fig:random_structure_prediction} have unit cells containing only 2 organic cations.
Fig.~\ref{fig:random_structure_prediction} also highlights which structures were present in the training set of the MLIP model. For the left-most structures, samples of these perovskites acquired from molecular dynamics during the active learning process are present in the training set of our model. For the next set of structures, the organic cation is present somewhere in the dataset, but the combination of cation and inorganic layer is not present. For the four right-most structures, the organic cation is not present in the dataset.

In all but three cases the identified structures with lowest energy correspond to the energy of the experimental structure. Subsequent comparison showed that these structures did indeed match the experimentally reported version. Therefore, the combination of a simple random structure searching scheme with the developed MLIP can successfully identify the ground state structure of these complex systems.

In the three cases for which the lowest energy structure does not match the experimentally reported structure, one structure search failed to find any structures with energy as low as that of the experimental structure  within the 200 searches (the lowest energy found was about 2 meV/atom higher than the energy of the experimental structure). In the other cases, we confirmed that the procedure found the experimental minima as well as a lower energy structure. Subsequent evaluation with DFT revealed that these structures were also assigned a lower energy than the experimental structure by DFT. 

\subsection{Prediction and Synthesis of new 2D Hybrid Perovskites}
\label{sec:experimental}

Finally, the structure search process was performed for a new organic cation with no previously known corresponding perovskite. Specifically, the combination of \textit{cis}-1,3-cyclohexanediamine with a Pb--I inorganic layer was studied. This molecule is not present in our dataset, but consists of chemical groups which are well represented. 

The structure searching procedure was conducted with a unit cell containing 8 copies of the organic cation, across two layers. In total, 6000 samples were generated, with the large number being required due to the large number of molecules present in the unit cell. The perovskite was synthesised via slow hydrothermal growth and the resulting structure was determined, at 200~K, using a diffractometer as described in section \ref{sec:methods_experimental}. 

\begin{figure*}
    \centering
    \includegraphics[width=1.0\textwidth]{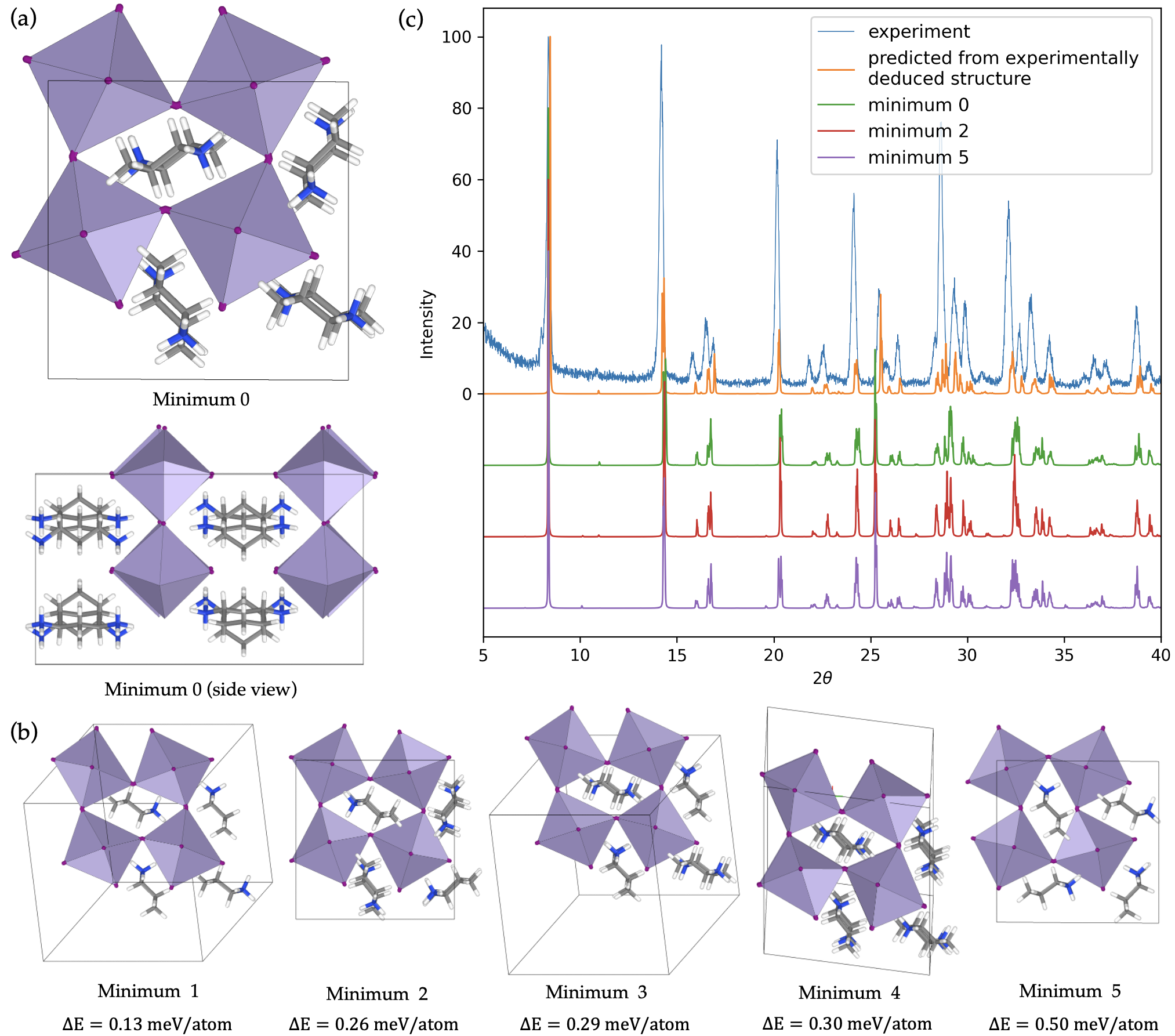}
    \caption{Comparing the lowest energy structures during the structure prediction task for a \textit{cis}-1,3-cyclohexanediamine based perovskite. Since these unit cells are relatively complex, and differences between structures are subtle, we have tried to find `equivalent' representations of the unit cells for comparison. Cif files of all structures are available. (a) The lowest energy minimum found during our procedure, which is equivalent to the structure deduced by experiment. The unit cell contains 8 molecules and 232 atoms. (b) The next 5 lowest lying minima, and their energy above the lowest structure. All the structures shown contain 8 molecules, however when a structure adopts a higher symmetry and hence a smaller unit cell, some molecules appear to hide behind others. The key difference between minimum 0 and minimum 4 is the out of plane lattice vector. (c) Comparison of the experimentally measured pXRD with some chosen minima. ``Experimentally deduced structure'' refers to the unit cell as deduced using the ShelXT program from data obtained by diffractometer (see also section \ref{sec:methods_experimental}).}
    \label{fig:synthesized_structure}
\end{figure*}

Figures \ref{fig:synthesized_structure}a and \ref{fig:synthesized_structure}b show the resulting lowest energy structure (denoted as ``minimum 0''), as well as the 5 next lowest energy structures that were predicted. As shown in the figure, the energy differences between the lowest lying minima are extremely small, with the 5 next best minima being only 0.5 meV/atom higher in energy the ground state. This energy difference is smaller than the likely error in our model, as well as the error of DFT due to finite k-point sampling. 

Several interesting points can be made about these results: firstly, the lowest lying minimum found by the structure search process agrees with the experimentally measured structure. Since our model is fitted to DFT data which does not perfectly match reality, differences are unavoidable in quantities such as equilibrium bond lengths, where the PBE functional makes an error. However, we can confirm that we predict the right structure by performing a geometry relaxation, with our model, of the experimentally reported structure. This resulted in exactly minimum 0, and the relaxation trajectory only involved only minor changes in bond length, as shown in the supplementary information.

The 5 next lowest energy minima all involve similar orientations of the organic cation, but differ in the set of reflections applied to the cations or the out of plane stacking vector. It is interesting to examine how easily one could differentiate between these structures using different experimental techniques. This was done by measuring the powder x-ray diffraction pattern (pXRD) of the synthesised perovksite, and comparing to the computed pattern of the lowest energy minima, as shown in Fig. \ref{fig:synthesized_structure}c. We compare the experimental result to that predicted from the experimentally reported structure, as well as minima 0, 2 and 5. The differences between the predicted XRD of the experimentally reported structure and minimum 0 (orange and green in Fig. \ref{fig:synthesized_structure}c) come only from the aforementioned small differences in bond lengths. Interestingly, the spectra of the three numbered minima are almost indistinguishable; it would be extremely difficult to robustly differentiate these structures from the pXRD alone. 

Furthermore, the small differences in molecular stacking lead to different optical properties. For example, out of the six structures in Fig. \ref{fig:synthesized_structure}, only minimum 0 is centrosymmetric. This means that it will not exhibit circular dichroism which is necessary for certain applications of 2D perovskites. When targeting certain properties, a full picture of the landscape of low energy minimum is clearly important. Our structure searching method offers a window into this landscape, which could be used to choose experimental methods, or gain confidence in conclusions.

Further predictions were made for 4 other organic cations which had no previously known perovskite structures. These are discussed in the supplementary information.

\subsection{Scalability and Computational Cost}

Performing the above process requires geometry relaxation of many large crystal structures, starting from high energy configurations. Typically, several hundred relaxations are required with hundreds to thousands of dynamics steps for each relaxation.

In the structure searching process for the 13 structures in Fig.~\ref{fig:random_structure_prediction}, the average unit cell size was 78 atoms and 200 samples were generated for each system. The entire set of calculations used to produce Fig.~\ref{fig:random_structure_prediction} was performed in only 20 hours on a single A100 GPU. This suggests that wide searches can be performed using modest computational resources. By comparison, a {\em single} DFT relaxation of one sample of the randomly generated structures shown in Fig.~\ref{fig:random_relaxation_example} (similar in size and complexity to Fig~\ref{fig:random_structure_prediction} structures), performed on two nodes (256 cores) of AMD EPYC 7742, can take more than one day. Furthermore, the cubic scaling of the DFT calculation makes the same task for much larger systems infeasible. 

One can also see the computational advantage of using our model in the structure search for the newly synthesised perovskite (section \ref{sec:experimental}), which has 8 molecules and 232 atoms in the unit cell. For each sample of the 6000 generated structures, relaxation took between 2000 to 4000 steps, leading to a total computational cost of 240 GPU hours. We estimate that performing the same relaxations with DFT would require approximately 1.2 million CPU node-hours. In this case the speedup corresponds to a factor of $10^4$. Note that the absolute times of course depend on the type of GPU and CPU hardware making a direct comparison not straightforward, but in terms of costing computational resources, an A100 GPU hour is approximately comparable to a node-hour with 128 CPU cores, and hence is the basis for the figures given above.

\section{Extrapolation to Underrepresented Organics}

\begin{figure}
    \centering
\includegraphics{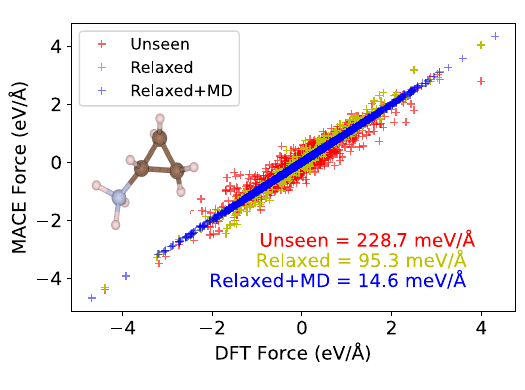}
    \caption{The force parity plot for cyclopropanaminium for three differently trained potentials. An `unseen' MLIP, with no samples of cyclopropanaminium, relaxed-model which has 200 randomly selected samples from the relaxation trajectories of the structure prediction model in the training set, and relaxed+MD which takes the top 10 most stable structures, followed by samples taken uniformly from MD trajectories. The two former MLIPs were trained independently.}
    \label{fig:c3}
\end{figure}

We have demonstrated good performance of the trained MLIP both in terms of single point accuracy and in relaxing randomly generated HOIPs structures to global minima. However, it is the case that many of the organic molecules in the test set are structurally and chemically similar to the organics in the training set. Here we demonstrate an example of organic cation in our test set, cyclopropanaminium (shown in Fig.~\ref{fig:c3}), for which the model performs poorly and suggest an efficient way retraining the model to improve the predictions.

For the original MLIP, which has not been trained on any examples of cyclopropanaminium, committee MD simulations immediately exceed the prescribed uncertainty threshold, indicating an uncertainty of the model in predicting forces. On samples that are taken from this MD, the model makes a large error with respect to DFT with a force RMSE of 228.7 meV/\AA. 

One approach would be to add some of these uncertain high-error samples to the training set and use the AL cycle to improve the potential for that specific organic. This is not possible when no experimental structure is known, since an initial structure is needed for running the active learning. As shown in the supplementary information Section II, we tested several approaches in which DFT calculations of only the isolated organic molecule were added to the training set, but these failed to improve the accuracy of the model to an acceptable level. 

Another way of approaching this problem is to use the structure prediction algorithm to generate HOIP structures with the new organic cation. One can relax these candidates with the model, take samples from the relaxation trajectories, and add them to the training set. As shown in Fig.~\ref{fig:c3}, when the model is trained with 200 distinct samples from relaxation trajectories of cyclopropanaminium lead iodide, the resulting force RMSE is 95.3 meV/\AA, a slight improvement over the original model. The meager improvement may be because many of the predicted relaxed structures are very similar, in terms of bond distances and orientation of the organic and inorganic components. A successful approach is to combine the structure prediction tool with MD simulations. Instead of taking 200 samples from the relaxations trajectories, we take only the 10 most stable structures predicted by the structure prediction algorithm, run short MD simulations (5 ps) and take samples uniformly every 1 ps from the MD trajectories. Note that we do not terminate the MD simulations based on uncertainty. Using this to add new data, the error in forces drops to 14.6 meV/\AA, within the range of previously seen compositions in the training set. This is achieved with only a total of 50 new samples, and in one cycle of retraining.

This approach works because the original MLIP predicts physically reasonable structures, despite the large error in forces with respect to DFT. In particular, the model relaxes the randomly generated configurations to sensible structures, in terms of the organic-inorganic stacking pattern. Similarly, MD simulations, while they may exhibit high committee uncertainty in forces, do not lead to unrealistic structures (no bond-breaking or coalescence of atoms). 

\section{Conclusions}

We have presented an efficient, accurate and general machine learning force field (MLIP) using the MACE architecture for lead based 2D HOIPs involving organic cations containing carbon, hydrogen, nitrogen and oxygen. Our model performs well on single point energy and force predictions on samples taken from molecular dynamics simulations and can extrapolate to unseen organic cations. 

Furthermore, the model is appropriate for performing high throughput screening of this class of materials. A simple random structure search procedure has been presented which, when paired with the MACE model, is able to rediscover the experimentally reported structure for a number of 2D perovskites in our database. The model is demonstrably accurate during this process, correctly reproducing the complex energy landscape, as shown by exploring specific examples with DFT. The computational cost of our structure generation process and model is small enough that this procedure can be applied at scale. 

Finally, our method was validated by synthesising a new perovskite composition. Besides predicting the correct structure, the model revealed a delicate landscape of low-lying energy minima, which in its self could be a useful investigative capability.

\section{Data Availablility Statement}

The committee of MACE models trained on the full training set is available in a zenodo repository \url{10.5281/zenodo.10729400}. The full train and test sets are also available as a python pandas dataframe. The experimentally determined newly synthesized structure, as well as the five predicted lowest energy structures found by our process, are also available. The random structure generation algorithm was implemented in a python package which can be found at \url{https://github.com/WillBaldwin0/LDHP-builder}.

\section{Acknowledgements}

WB, CS, and CG thank the AFRL for partial funding of this project through grant FA8655-21-1-7010.
CS and NK gratefully acknowledge the University of South Carolina for the support provided through start-up funds and  support from the U.S. Department of Energy, Office of Science, Basic Energy Sciences (DE-SC0022247). This work utilized computational resources from the ARCHER2 UK National Supercomputing Service (http://www.archer2.ac.uk) which is funded by EPSRC via the membership of the UK Car-Parrinello Consortium, the Cambridge Service for Data Driven Discovery (CSD3), and University of South Carolina Hyperion HPC cluster. The authors thank Volker Blum for discussions about the utility of structure searching for these materials. 

\section{Conflict of Interest}
CG is director of Symmetric Group LLP, which markets some MACE models commercially.  The other authors declare that they have no conflicts of interest. 

\section{Methods}

\subsection{MACE Machine Learning Interatomic Potentials}

This work has utilised the MACE framework for constructing machine learned interatomic potentials \cite{mace}. MACE is a recently developed equivariant message passing tensor network which offers state of the art accuracy. The MACE architecture has been described and evaluated in detail previously \cite{MACE2022, KovacsMACEeval2023, maceoff23, batatia2024foundation}. Therefore, the following description simply discusses some key aspects of the model design. 

A MACE model predicts the total energy of a system as a sum of atom centered contributions. The environment around an atom is described by the atomic number and relative positions of neighbouring atoms, up to some fixed cutoff: $\mathcal{N}(i) = \{z_j, \mathbf{r}_{ij}\}_{j | r_{ij} < r_{cut}}$. The MACE architecture utilises ideas from the atomic cluster expansion to efficiently construct atom centered features based on the local environment. These atom centered features are many-body, in that they depend simultaneously on atomic numbers and positions of several neighbours in a non-trivial way. These features are iteratively updated, and the final energetic contribution from each atom is expressed as a learnable function of these features.

The specifications of the MACE models used in this work are, in the nomenclature of reference \cite{maceoff23}, given in table \ref{table:model_specs}.

\begin{table}
\centering
\caption{Specifications of MACE models used in this study}
\begin{tabular}{p{5cm}c}
\hline
number of chemical channels & 128 \\
\hline
maximum equivariance order L & 1 \\
\hline
single layer cutoff radius (\AA) & 5  \\
\hline
number of layers & 2  \\
\hline
\end{tabular}
\label{table:model_specs}
\end{table}

\subsection{Molecular-Dynamics and Geometry Relaxations with MACE potentials}
\label{sec:methods_md}

All molecular dynamics (MD) simulations were carried using the atomic simulation environment (ASE) package \cite{ase} in the NPT ensemble at 300 K and 1 atmosphere. A Nosé–Hoover thermostat \cite{nose,hoover} was used. During active learning, MD simulations were propagated using the average prediction of 3 committee members. The relative force uncertainty $f_{rel}^i$ is defined as
\begin{equation}
f_{rel}^i=\frac{\sigma_i}{\left|\bar{F}_i\right|+\epsilon},
\label{eq:rel_force_uncert}
\end{equation}
where $\sigma_i$ and $\bar{F}_i$ denote the standard deviation and mean of forces over the committee members on atom \textit{i}. $\epsilon$ is a regularizer to avoid diverging ratios for small forces. At each MD step the atom with the greatest $f_{rel}^i$ is selected, and this value is compared against the predefined threshold of 0.2. If this uncertainty indicator exceeds the threshold, the simulation is terminated. The regularizer $\epsilon$ for all the simulations was set to 0.2 eV/\AA.

All geometry relaxations have been done using preconditioned LBFGS as the optimiser\cite{preLBFGS}. During relaxations, both cell sizes and atomic positions are allowed to change and the relaxed cell is achieved when the maximum force on each atom is less than 1 meV/\AA.

\subsection{Electronic Structure Calculations}

All the electronic structure calculations for either relaxation or single point calculations are performed using Vienna Ab initio Simulations Package (VASP) \cite{vasp1,vasp2} with the PBE \cite{pbe} for the exchange-correlation functional and the projector augmented-wave (PAW) pseudopotentials \cite{paw1,paw2}. Dispersion energy-corrections are applied using D3 approximation \cite{d3}. A $\Gamma$-centered Monkhorst-Pack \cite{MP} k-point grids are used to sample the Brillouin zones, with a density of 1000 k-points per number of atoms, with divisions along each reciprocal lattice vector proportional to its length, as implemented in pymatgen \cite{pymatgen}. The electronic wave functions were expanded in a plane wave basis set with an energy cutoff of 600 eV.  

\subsection{An Algorithm for Random Structure Generation of 2D HOIPs}
\label{section:methods_random_algo}

A random structure generation algorithm has been developed to demonstrate the usefulness of the MACE model. Our algorithm is not intended to be completely general and relies on several simplifications. Future developments could utilise methods from organic crystal structure prediction for more generality. The code used in this project is available as a python package at \url{https://github.com/WillBaldwin0/LDHP-builder}.

The procedure is as follows: The inorganic layer is first generated from lead and the chosen halide as a monolayer of regular lead--hailde octahedra. The lead--halide bond length is chosen to be the average of such bonds across our training set. 

For each organic cation to be placed into the unit cell, the following process is performed. We assume that the molecule joins to the layer in a certain way: Salient points on the molecule are defined as the heavy atoms on the `extremities'. In practice this is done by first finding the moment of inertia tensor of the molecule and interpreting the eigenvectors as a local coordinate basis for the molecule. For molecules which are longest in a certain direction, the eigenvector with the smallest eigenvalue is generally directed along this direction. The extremities of the molecule are defined as the heavy atoms which have the largest relative distance between one another when projected onto this vector. One of these heavy atoms serves as a reference atom which is placed onto a given point on the inorganic layer. The orientation of the cation is then determined by first applying a random rotation about the selected atom, and subsequently applying up to two reflections in planes normal to the lattice vectors of the inorganic monolayer. 

After repeating the above for each molecule in the unit cell, the molecules are checked for intersections. Assuming there are no intersections, the out of plane lattice constant is fixed to remove as much vacuum from the cell as possible. The result of this procedure is a monolayer with a set of organic cations at a random orientation on the layer. See Fig. \ref{fig:random_relaxation_example}b for example structures from this procedure. 

\subsection{Synthesis and Characterization of Perovskite Materials for Verification of Modeled Results}
\label{sec:methods_experimental}

The perovskite \ce{1,3-(\textit{cis})-cyclohexanediamine-PbI4} was synthesized in order to compare the observed crystal structure with the results obtained via the computational methods described previously. Crystals of the perovskite were obtained through slow hydrothermal growth by dissolving equimolar amounts of the amine and lead (II) iodide in concentrated hydriodic acid in a sealed pressure vessel at 150 $^{\circ}$C and cooled at a rate of 5 $^{\circ}$C per hour, resulting in the formation of mm-scale orange crystalline chunks. Residual hydriodic acid was removed by washing with methylene chloride and diethyl ether, followed by drying under vacuum for several days. The crystal pieces are highly stable to ambient atmosphere and demonstrate no signs of decomposition over weeks of storage.

The crystal structure was determined using a Rigaku XtaLAB Synergy diffractometer. Crystal samples were mounted in oil on a ring-loop and placed in a cryo N$_2$ stream at 200 K. CrysAlis Pro was used to screen and collect diffraction patterns using Mo K$_{\alpha}$ ($\lambda= 0.71073$ \AA). A full sphere of diffraction data was collected, and multiscan empirical absorption correction was applied. The maximum resolution that was achieved was $\Theta$ = 31.00$^{\circ}$ (0.69 \AA). The structures were solved with the ShelXT (Sheldrick, 2016) structure solution program using the Intrinsic Phasing solution method and by using \textbf{Olex2} (Dolomanov et al., 2009) as the graphical interface. The model was refined with version 2016/6 of ShelXL 2016/6 (Sheldrick, 2015) using Least Squares minimisation. The crystal structure was determined with minimal guidance beyond initial atomic assignment and the resulting solved structure featured a low R value indicating that the solved structure aligned well with the atomic positions observed in the diffraction pattern.

Powder XRD was measured on a Rigaku SmartLab as an additional point of comparison between both the predicted and experimental crystal structures to assess the presence of any additional crystal phases at room temperature that may contribute to different structural behavior. Samples were prepared from the as-grown \ce{ABX4} perovskite crystals by grinding in a mortar and pestle to ensure uniform distribution of powder particle size and orientation. All measurements were performed at room temperature under ambient atmosphere. The $\theta / 2\theta$ spectra of the perovskite powders were then compared to predicted patterns generated from either the experimental or as-modeled crystal structures. 

Simulations of pXRD were performed in the VESTA structure visualization software package \cite{vesta}. 

\bibliographystyle{IEEEtran}
\bibliography{hoipbib}

\end{document}


\title{Supporting Information: \\ Accurate Crystal Structure Prediction of New 2D Hybrid Organic Inorganic Perovskites}

\author{Nima Karimitari}
\thanks{These authors contributed equally}
\affiliation{Department of Chemistry and Biochemistry, University of South Carolina, South Carolina 29208, United States}

\author{William J. Baldwin}
\email{wjb48@cam.ac.uk}
\affiliation{Department of Engineering, University of Cambridge, Cambridge CB2 1PZ, UK}

\author{\hspace{-4pt}$^{,\hspace{2pt}\ast}$ \text{Evan W. Muller}}
\affiliation{UES, Inc., Beavercreek, Ohio 45432, United States}

\author{Zachary J. L. Bare}
\affiliation{Department of Chemistry and Biochemistry, University of South Carolina, South Carolina 
29208, United States}

\author{W. Joshua Kennedy}
\affiliation{Materials and Manufacturing Directorate, Air Force Research Laboratory, Wright-Patterson AFB, Dayton, Ohio 45433, United States}

\author{G\'{a}bor Cs\'{a}nyi}
\affiliation{Department of Engineering, University of Cambridge, Cambridge CB2 1PZ, UK}

\author{Christopher Sutton}
\email{cs113@mailbox.sc.edu}

\affiliation{Department of Chemistry and Biochemistry, University of South Carolina, South Carolina 29208, United States}

\date{\today}

\maketitle

\section{Predictions for the Newly Synthesised Perovskite}

\begin{figure*}
    \centering
    \includegraphics[width=0.7\textwidth]{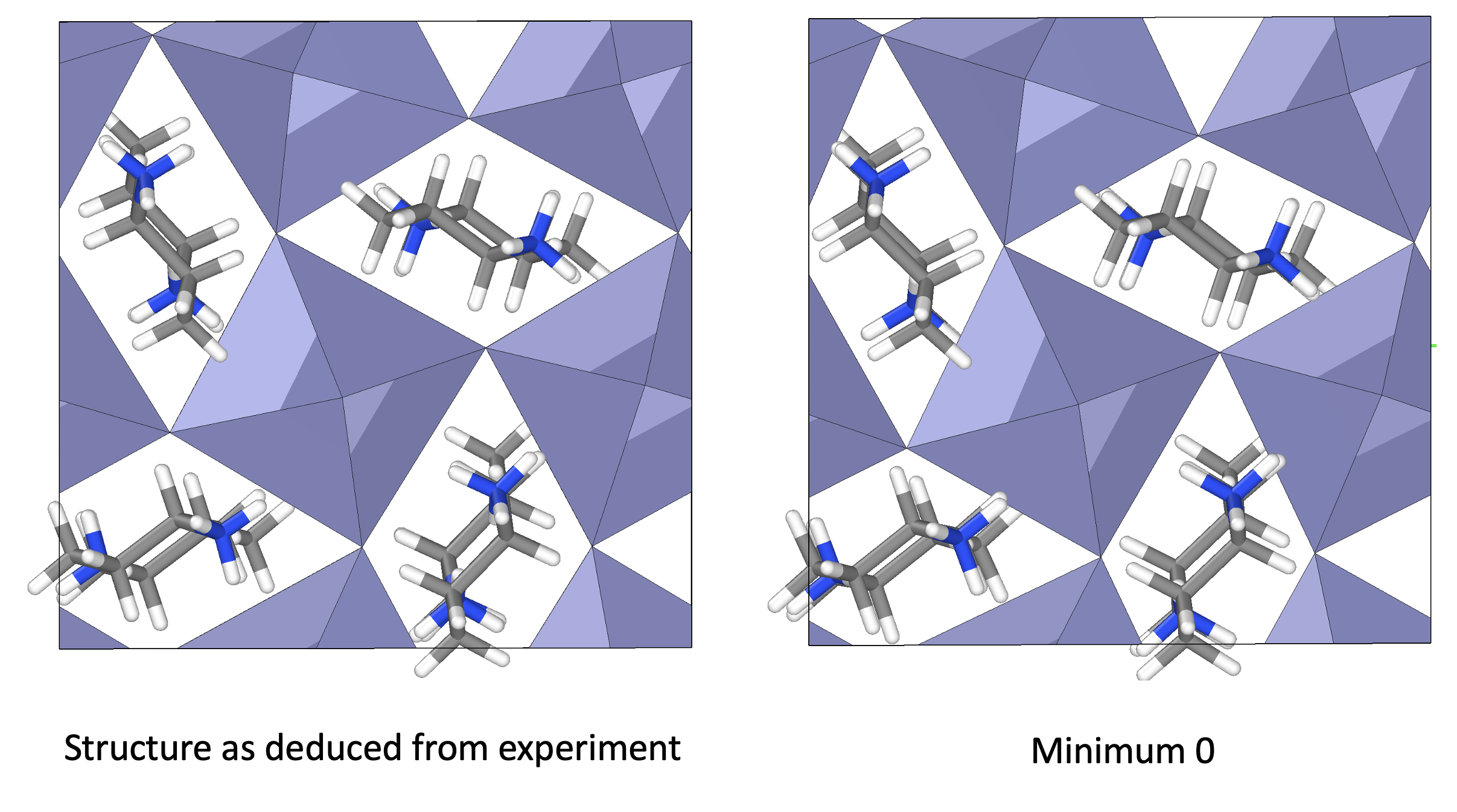}
    \caption{Comparing the structure as deduced from experiment to the lowest minima found by our process.}
    \label{fig:exp_comp}
\end{figure*}

When validating our prediction of the newly synthesised perovskite, numerical comparisons are similar but do not perfectly match since our model is trained to reproduce the energy landscape of DFT calculations with PBE+D3 functional. Therefore, there is an overestimation of 3-5\% in bond length distances compared to the experiment, as expected from PBE+D3 functionals. We confirm our prediction is correct by firstly examining the two structures by eye, which are shown in Fig. \ref{fig:exp_comp}. Secondly, we showed that a geometry relaxation of the experimentally reported structure with our model results in exactly the predicted structure. Even though the figure shows that this does not meaningfully change the structure, this is the origin of the differences between the predicted pXRD of the experimentally deduced structure and our prediction in the main text Fig. 8. 

The computational cost of the structure prediction algorithm is highly dependent on the number of molecules in the unit cell since this dictates the number of samples required. In our dataset, the median number of molecules per unit cell (for perovskites with no locally 3D regions) was 4. Evidently there is a trade-off between running fewer, expensive, searches with large unit cells, or many cheaper searches with small cells and accepting that one will occasionally miss the correct structure.

\section{Extrapolation to 1D Perovskites}

While conducting random structure searches of potential new HOIPs, it was discovered that the final MACE model is able to extrapolate well to 1D HOIPs, without having seen any in the training set. 

\begin{figure*}
    \centering
    \includegraphics[width=0.7\textwidth]{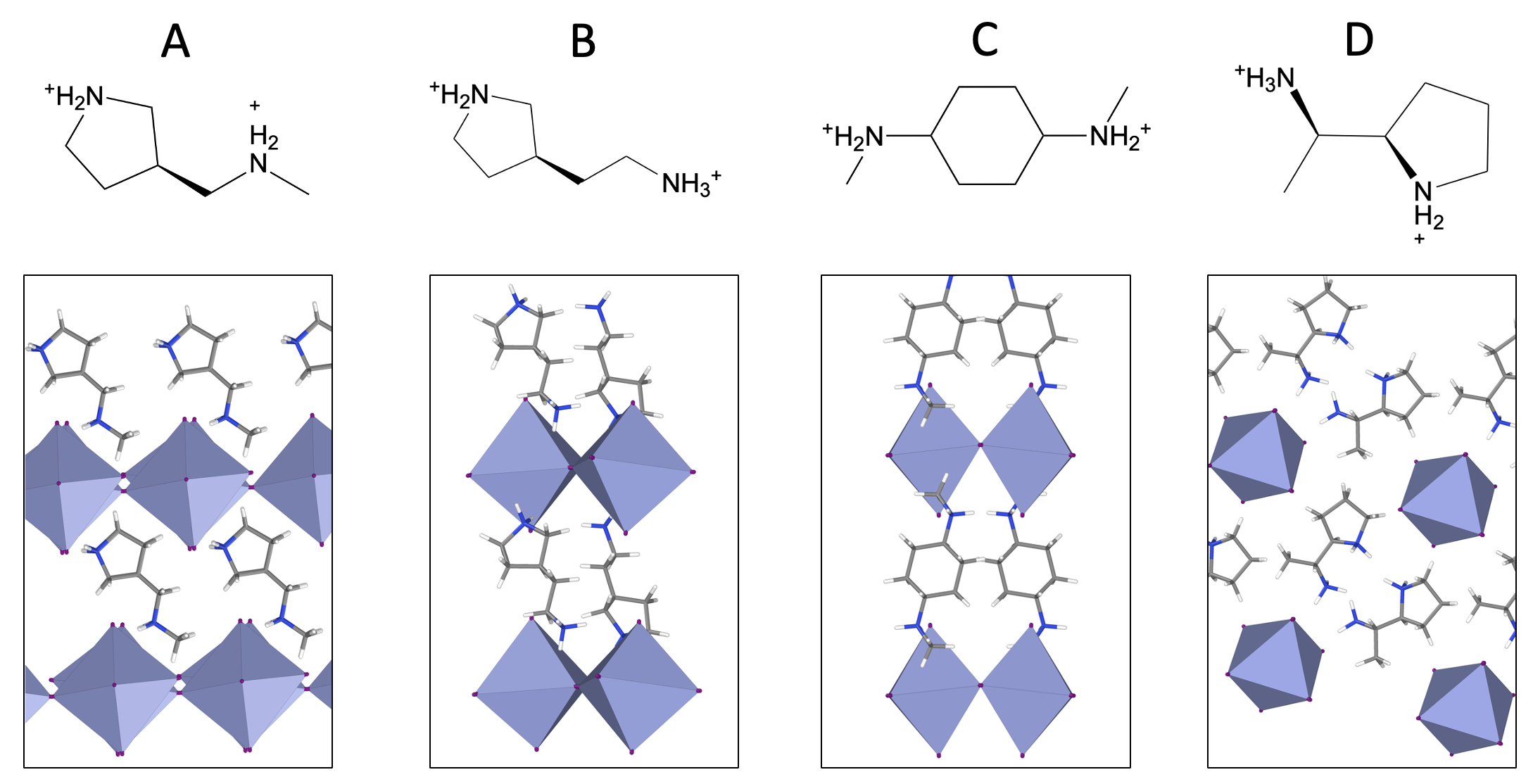}
    \caption{Applying our random structure search procedure to 4 molecules with chiral centers. The lowest energy structure found for the right-most molecule relaxed to a 1D perovskite.}
    \label{fig:chrial_organics}
\end{figure*}

\begin{figure}
    \centering
    \includegraphics[width=.5\textwidth]{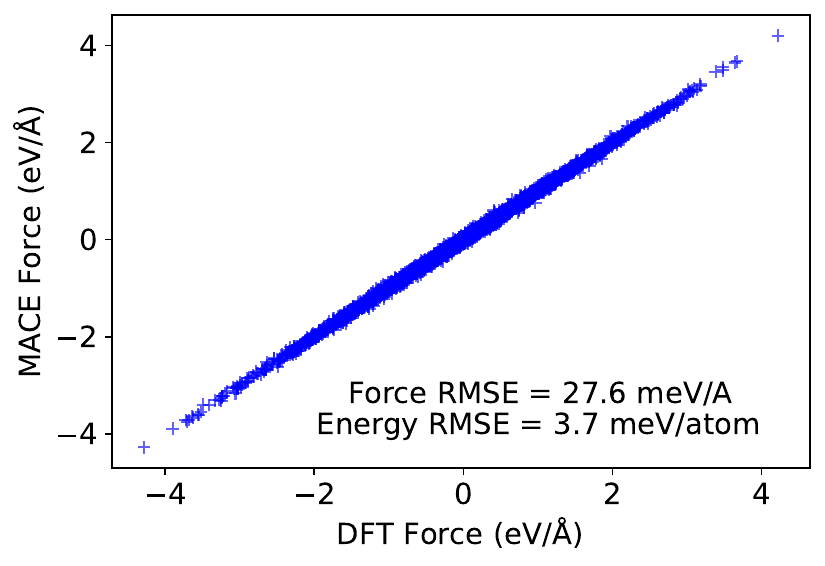}
    \caption{The force parity plot between the model and DFT on samples taken from an MD simulations of predicted 1D HOIP.}
    \label{fig:parity}
\end{figure}

We applied the random structure search procedure to four organic cations with chiral centers, as shown in Fig.~\ref{fig:chrial_organics} and predicted the most stable HOIPs that they can form. These structure searches were performed with a small unit cell, containing just 2 molecules and one inorganic layer. An interesting case is molecule D in Fig.~\ref{fig:chrial_organics}, 2-(1-aminoethyl)pyrrolidinium, for which the most stable structure is a 1D HOIP. This is surprising because the structure search procedure was the same as described in the main text, where the initial generated samples are all 2D layered \ce{PbX4} structures, and the model was able to achieve a structure with \ce{PbX3} inorganic unit (the additional halide ions are isolated, away from the rest of the inorganic framework). To examine the accuracy of the model and stability of this structure, a constant pressure MD simulation of the structure was ran at 300 K, 1 atm for 50 ps. The maximum relative force uncertainty remains below the threshold (as discussed in main text section II) at all times. Furthermore, the root mean square error with respect to DFT (see also Fig.~\ref{fig:parity}), for samples taken every 5 ps was 27.5 meV/\AA and 3.7 meV/atom for forces and energy, respectively. This close agreement with DFT implies that the current trained model, which has not trained to any lower dimensional HOIPs, is accurate in dealing with 1D HOIPs. 

\section{Underrepresented organics}

One of the challenges of the trained potential is in dealing with organic molecules with local structural features and functional groups that are very different to those found in the training dataset. One example, as discussed in the main text Section V is cyclopropanaminium. Several other approaches were tested while searching for ways to quickly add new information about unseen organic molecules. Some of these methods all involved performing calculations of isolated molecules, and adding just these to the training set. Here we summarise out the outcomes of these analyses.

\begin{figure*}
    \centering
    \includegraphics[width=1.0\textwidth]{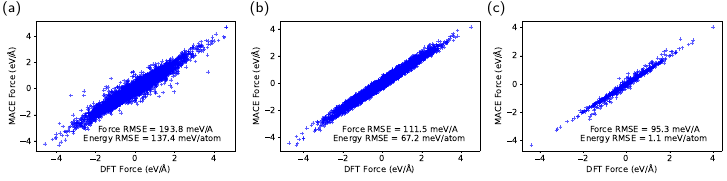}
    \caption{Force parity plots for three different analysis of extrapolating to underrepresented molecules. All are taken from independent MD simulations with (a) trained to isolated molecules in box, (b) trained to only the forces of the isolated molecules, and (c) trained to the samples taken from the relaxation trajectory.  }
    \label{fig:c3_si}
\end{figure*}

\begin{enumerate}
    \item[(a)] \textbf{Isolated Molecules.} We place the isolated molecule in an empty box ($15 \times 15 \times\ 15$ \AA) and ran collected samples by running molecular dynamics simulations using extended tight binding (xTB)\cite{xtb}. Specifically, the GFN2-xTB\cite{GFN2XTB} was used via the xtb-python api. Simulations were ran at 500K using a time step of 1 fs and 200 samples were taken at intervals of 1 ps. These configurations were subsequently evaluated with DFT using the same settings as the perovskite training data. One difficulty with this method is that cations in HOIPs are charged molecules and ignoring the charge in the isolated molecule calculations can severally affect the force/energy errors. However, including them is also a problem due to long range interactions between neighboring unitcells. We considered the charged cations but with the dipole and quadropole corrections \cite{long_range}. The original model retrained to this dataset shows higher than expect errors as shown in Fig.~\ref{fig:c3_si}a. The retrained model have a very poor prediction for energies. This is because even with the presence of corrections to long-range interactions, the energy difference between charged and uncharged calculations is so large that the current MLIP, trained to cations that are present in the periodic systems, cannot accurately predict both isolated molecules and molecules in solids.
    \item[(b)] \textbf{Isolated Molecules, Force Information Only.} As a final test on the isolated molecules, we then tried to train the original model to only forces of the isolated cyclopropanaminium. This modification led to significant improvement in both force and energy predictions, as shown in Fig.~\ref{fig:c3_si}b, but still the error was an order of magnitude larger than the values reported for seen perovskites in Section II.B.
    \item[(c)] \textbf{Structures found by Random Structure Search.} Instead of dealing with the isolated molecules, we used the random structure search procedure to generate initial HOIPs with this molecule and then relax them with the MLIP. While the MLIP used in the structure search struggles with energy and forces, it can still create samples that are useful for retraining the model. In Fig.~\ref{fig:c3_si}c, we retrained the original model to 200 randomly selected samples from the relaxation trajectories. This leads to an improvement with errors of 95.3 meV/\AA and 1.1 meV/atom for forces and energy.
    \item[(d)] \textbf{Random Structure Search + MD.} In our final experiment, we combined the previous step with MD simulations. This is because in relaxation trajectories, many of the samples are structurally similar making the learning inefficient, but MD simulations can lead to more diverse local environments that the model can learn from. Therefore, we collected the 5 most stable structures predicted by the search algorithm method, ran MD simulations for 10 ps. In total 50 samples (taken every 1 ps) are then added to the original training set and the model is retrained. The errors are at 14.6 meV/\AA and 0.9 meV/atom also shown in main text in Sec. V. 
\end{enumerate}
\bibliographystyle{IEEEtran}
\bibliography{si}